\documentclass[superscriptaddress,twocolumn,showpacs,preprintnumbers,amsmath,amssymb,prl]{revtex4} 

\usepackage[dvipdfm]{graphicx}
\usepackage{dcolumn,upgreek}
\usepackage{bm}
\usepackage{color}

\def\ii{{\mathrm{i}}}
\def\ff{{\mathrm{f}}}
\def\ee{{\mathrm{e}}}
\def\dd{{\mathrm{d}}}

\def\no{{\nonumber}} 

\def\ket#1{|#1\rangle}

\def\bracketii#1#2#3{\langle #1 | #2| #3\rangle}

\def\sub#1{_\mathrm{#1}} 

\begin{document}


\title{Classical Realization of Dispersion Cancellation by Time-Reversal Method}

\author{Kazuhisa Ogawa}
\email{ogawa@giga.kuee.kyoto-u.ac.jp}
\affiliation{%
Department of Electronic Science and Engineering, Kyoto University, Kyoto 615-8510, Japan
}%
 
\author{Shuhei Tamate}
\affiliation{%
RIKEN Center for Emergent Matter Science, Wako-shi, Saitama 351-0198, Japan
}%
\author{Toshihiro Nakanishi}
\affiliation{%
Department of Electronic Science and Engineering, Kyoto University, Kyoto 615-8510, Japan
}%
\author{Hirokazu Kobayashi}
\affiliation{%
Department of Electronic and Photonic System Engineering,\\
Kochi University of Technology, Tosayamada-cho, Kochi 782-8502, Japan
}%
\author{Masao Kitano}
\affiliation{%
Department of Electronic Science and Engineering, Kyoto University, Kyoto 615-8510, Japan
}%

\date{\today}

\begin{abstract}
We propose a classical optical interferometry scheme that reproduces dispersion-insensitive Hong--Ou--Mandel interferograms.
The interferometric scheme is derived from a systematic method based on the time-reversal symmetry of quantum mechanics. 
The scheme uses a time-reversed version of a Hong--Ou--Mandel interferometer with pairs of orthogonally polarized input laser pulses.
We experimentally demonstrate automatic dispersion cancellation using the interferometry.
The results show that the interferometer can obtain high-visibility interferograms with high signal conversion efficiency.
\end{abstract}

\maketitle

Hong--Ou--Mandel (HOM) interference \cite{PhysRevLett.59.2044} is a fundamental two-photon interference phenomenon in quantum optics.
The study of HOM interference has inspired a variety of studies on two-photon interference phenomena, such as quantum beating \cite{PhysRevLett.61.54}, phase superresolution \cite{PhysRevA.42.2957,PhysRevLett.65.1348,PhysRevLett.89.213601,walther2004broglie}, and phase supersensitivity \cite{nagata2007beating,okamoto2008beating}. 
Due to the time-frequency correlation of photon pairs, HOM interferograms are insensitive to even-order dispersion.
This property is utilized in automatic dispersion cancellation \cite{PhysRevA.45.6659,PhysRevLett.68.2421,PhysRevA.88.043845} and has been applied to precise measurements of length or time in dispersive media such as biological samples and optical fibers.
These applications include quantum-optical coherence tomography \cite{PhysRevA.65.053817,PhysRevLett.91.083601,nasr2009quantum}, quantum clock synchronization \cite{PhysRevLett.87.117902}, and the measurement of photon tunneling time \cite{PhysRevLett.71.708}.

Automatic dispersion cancellation in HOM interferometers often suffers from weak output signals due to low efficiency in the generation and detection of entangled photon pairs.
To address this, several schemes have been proposed to  achieve automatic dispersion cancellation with intense classical input light \cite{banaszek2007blind,PhysRevA.74.041601,resch2007classial,le2010experimental,kaltenbaek2008quantum,resch2009}.
One of the most successful schemes is chirped-pulse interferometry (CPI) developed by Kaltenbaek {\it et al.}~\cite{kaltenbaek2008quantum,resch2009,PhysRevLett.102.243601}.
CPI is implemented by using intense chirped and antichirped laser pulses as input light.
The interference signals are extracted from narrow-band light after sum-frequency generation (SFG), which is used as an efficient two-photon detector for optical coherence tomography \cite{Pe2007broadband}.
CPI can reproduce dispersion-insensitive HOM interferograms with much stronger signals.
Owing to the simplicity of the configuration, CPI has been applied to dispersion-insensitive optical coherence tomography \cite{Lavoie2009,mazurek2013dispersion}.

CPI is based on the {\it time-reversal method} \cite{PhysRevLett.98.223601,PhysRevA.88.063813},
which is a technique to reproduce the interferograms of an optical system by temporally reversing the original system. 
Resch {\it et al.}~first proposed the time-reversal method for reproducing multi-photon interferograms without an entangled-photon resource, and applied it to realizing phase superresolution \cite{PhysRevLett.98.223601}.
In their setup, however, the input light must be attenuated to photon counting level.
When applying the time-reversal method to realize automatic dispersion cancellation in practical applications, it is necessary to apply an additional technique to replace the attenuated input light with intense coherence light. 
In the case of CPI, by using chirped and antichirped laser pulses as input light, intense input light can be used in the time-reversed system.

In this Letter, we introduce a systematic method for constructing a time-reversed system reproducing dispersion-insensitive HOM interferograms with an intense classical light source.
Our scheme consists of the two parts. 
First, we construct a time-reversed system with a two-photon input state in accordance with the time-reversal method by Resch {\it et al.}~\cite{PhysRevLett.98.223601}.
Second, we modify the system to replace the two-photon input state with intense coherent input light.
Our interferometric scheme employs pairs of orthogonally polarized input laser pulses instead of the chirped and antichirped laser pulses used in CPI.
We also experimentally demonstrate automatic dispersion cancellation using this interferometry.
Our interferometry can achieve high-visibility interference with high efficiency owing to its simplicity.


\begin{figure}[t]
  \centering
  \includegraphics[width=8.5cm]{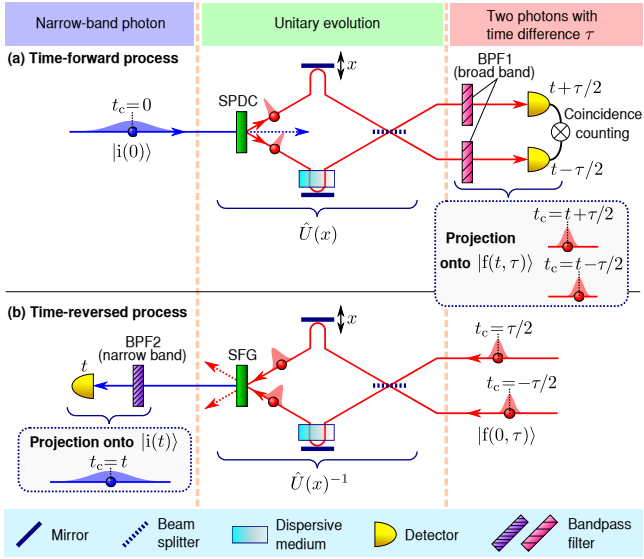}
\caption{(color online).
(a) Time-forward and (b) time-reversed HOM interferometer with time-frequency-entangled photon pairs. 
BPF1 and BPF2 are bandpass filters with relatively broad and narrow transmission spectra, respectively. 
The center frequency of the transmission spectrum of BPF2 is twice as large as that of BPF1.
} 
\label{Qthoery}
  \end{figure}

We start by introducing the theory of the time-reversal method.
The time-reversal method is based on the following identity for a unitary operator $\hat{U}$ and a pair of states $\ket{\ii}$ and $\ket{\ff}$:
\begin{align}
|\bracketii{\ff}{\hat{U}}{\ii}|^2=|\bracketii{\ii}{\hat{U}^{-1}}{\ff}|^2.
\label{eq:1}
\end{align}
The left-hand side of Eq.~(\ref{eq:1}) corresponds to the success probability of the {\it time-forward process}, where the initial state $\ket{\ii}$ evolves with $\hat{U}$ and is projected onto the final state $\ket{\ff}$. 
The right-hand side corresponds to the {\it time-reversed process}, where the roles of $\ket{\ii}$ and $\ket{\ff}$ are swapped and the unitary evolution $\hat{U}$ is inverted.
Thus the time-reversed process reproduces the same success probability as that of the time-forward process.

In the following, we apply the time-reversal method to an HOM interferometer and construct a time-reversed system with a two-photon input state reproducing dispersion-insensitive HOM interferograms.

The schematic setup of a time-forward HOM interferometer with time-frequency-entangled photon pairs is shown in Fig.~\ref{Qthoery}(a). 
The initial state is a narrow-band pump photon state $\ket{\ii(0)}$, represented by a wave function centered at $t\sub{c}=0$ in the time domain.
The input photon is converted into a time-frequency-entangled photon pair in a nonlinear crystal through spontaneous parametric down-conversion (SPDC).
One of the photons travels the upper arm, including a delay mirror with displacement $x$. The other travels the lower arm, including passing through a dispersive medium.
The two photons are combined at a beam splitter.
The overall time evolution is denoted by the unitary operator $\hat{U}(x)$.
 After passing through the bandpass filters BPF1s, the photon pair is detected in the upper and lower arms at time $t\pm\tau/2$, respectively. 
These detections are interpreted as the projection onto the final state $\ket{\ff(t,\tau)}$, which is represented by a pair of wave functions centered at $t\sub{c}=t\pm\tau/2$ in the time domain, respectively.
The widths of these wave functions are determined by the transmission spectra of the BPF1s.
The detection probability density is given by $|\bracketii{\ff(t,\tau)}{\hat{U}(x)}{\ii(0)}|^2$.
The success probability $P(x)$ of the coincidence counting is given by the integration of the detection probability density over $t$ and $\tau$:
\begin{align}
P(x)=\int^{\infty}_{-\infty}\dd t\int^{\infty}_{-\infty}\dd \tau|\bracketii{\ff(t,\tau)}{\hat{U}(x)}{\ii(0)}|^2.
\end{align}
By changing the displacement $x$, the success probability sharply drops only when the lengths of the two arms are balanced. 
This decrease in the success probability is called the HOM dip.
The width of the HOM dip is determined by the phase-matching condition of SPDC and the transmission spectrum of BPF1.

Based on the time-reversal method, the time-reversed HOM interferometer is constructed as shown in Fig.~\ref{Qthoery}(b).
The initial state is the two-photon state denoted by $\ket{\ff(0,\tau)}$, which is represented by a pair of wave functions centered at $t\sub{c}=\pm\tau/2$ in the time domain, respectively.
This state is the time-reversal counterpart of the final state in the time-forward process.
The pair of photons pass through the optical system in the reverse manner to the time-forward case and are converted into a single photon by SFG.
We finally detect the up-converted photon at time $t$ through the bandpass filter BPF2.
This detection is interpreted as the projection onto the final state $\ket{\ii(t)}$, represented by a wave function centered at $t\sub{c}=t$ in the time domain.
This state is a time-reversal counterpart of the initial state in the time-forward process. 
The detection probability density is given by $|\bracketii{\ii(t)}{\hat{U}(x)^{-1}}{\ff(0,\tau)}|^2$, which is equal to $|\bracketii{\ii(0)}{\hat{U}(x)^{-1}}{\ff(-t,\tau)}|^2$ due to the time-translation symmetry.
The success probability $P\sub{r}(x,\tau)$ is given by the integration of the detection probability density over only $t$, not $\tau$:
\begin{align}
P\sub{r}(x,\tau)&=\int^{\infty}_{-\infty}\dd t|\bracketii{\ii(0)}{\hat{U}(x)^{-1}}{\ff(-t,\tau)}|^2\no\\
&=\int^{\infty}_{-\infty}\dd t|\bracketii{\ii(0)}{\hat{U}(x)^{-1}}{\ff(t,\tau)}|^2.
\end{align}
By integrating $P\sub{r}(x,\tau)$ over all delays $\tau$, we can reproduce the same interferogram as the time-forward HOM interferometer $P(x)$. 
In our scheme, we measure each success probability $P\sub{r}(x,\tau)$ for various delays $\tau$ and sum the measured data over all delays $\tau$ by post-processing.
The width of the HOM dip is determined by the spectrum of the input pulse and the phase-matching condition of SFG.


\begin{figure}[t]
  \centering
  \includegraphics[width=8.5cm]{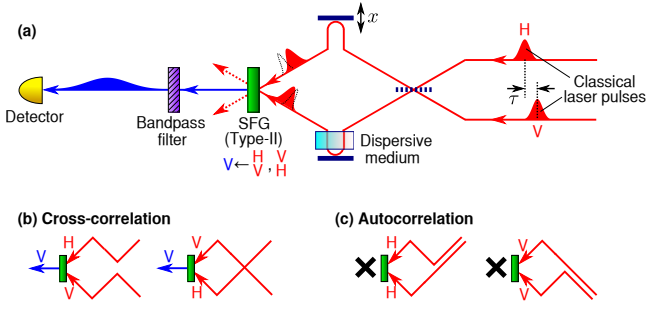}
\caption{(color online).
(a) Time-reversed HOM interferometer with intense coherent input light.
The labels H and V denote horizontal and vertical polarization, respectively. 
(b), (c) Feynman paths leading to SFG in this setup.
Type-II SFG extracts only pairs of orthogonally polarized photons (HV and VH) and thus eliminates the contribution from the autocorrelation terms in (c).
} 
\label{Ctheory}
  \end{figure}

We next consider the replacement of the two-photon input state with intense coherent light as shown in Fig.~\ref{Ctheory}(a). 
The time-reversed system with intense coherent input light has four Feynman paths leading to SFG, shown in Figs.~\ref{Ctheory}(b) and (c).
Since only the cross-correlation terms (b) lead to HOM interference, we have to eliminate the contribution from the autocorrelation terms (c).
In our system, we use pairs of orthogonally polarized (H and V) laser pulses as input light and employ a type-II nonlinear optical crystal for SFG.
Type-II SFG extracts only pairs of orthogonally polarized photons and thus eliminates the contribution from the autocorrelation terms (c).

For intense input light, we can calculate the interferogram of the time-reversed HOM interferometer by using only classical electromagnetics, as detailed below.
We denote the spectra of the two input pulses by $E(\omega)$ and $E(\omega)\ee^{\ii\omega\tau}$, and the effect of the dispersion medium is modeled by a linear transfer function $H(\omega)$.
The spectrum of mixed light through SFG, $E\sub{SFG}(\omega)$, is given by the following convolution integral:
\begin{align}
E\sub{SFG}(\omega)
\propto&\int^{\infty}_{-\infty}\dd\omega'E(\omega')E(\omega-\omega')H(\omega')\ee^{\ii(\omega-\omega')x/c}\no\\
&\hspace{2cm}\times\left[\ee^{\ii\omega'\tau}-\ee^{\ii(\omega-\omega')\tau}\right],
\end{align}
where $c$ is the speed of light in a vacuum.
Assuming that the transmission spectrum of the bandpass filter is sufficiently narrow and its center frequency is $2\omega_0$, the measured intensity $I\sub{r}(x,\tau)$ after the bandpass filter is given by $I\sub{r}(x,\tau)=|E\sub{SFG}(2\omega_0)|^2$.
By integrating $I\sub{r}(\tau,x)$ over all delays $\tau$, the post-processed interferogram $S(x):=\int^{\infty}_{-\infty}\dd \tau I\sub{r}(\tau,x)$ is calculated as
\begin{align}
S(x)\propto&\int^{\infty}_{-\infty}\dd\omega
|E(\omega_0+\omega)|^2|E(\omega_0-\omega)|^2|H(\omega_0+\omega)|^2\no\\
&-
\int^{\infty}_{-\infty}\dd\omega
|E(\omega_0+\omega)|^2|E(\omega_0-\omega)|^2\no\\
&\hspace{1.2cm}\times H(\omega_0+\omega)H(\omega_0-\omega)^*
\ee^{-\ii2\omega x/c}.
\end{align}
We can see that $S(x)$ is insensitive to even-order dispersion, which is represented as $H(\omega)=\ee^{\ii[\phi_0+\phi_2(\omega-\omega_0)^2+\cdots]}$.
For a Gaussian frequency spectrum $E(\omega)=\exp[-(\omega-\omega_0)^2/(2\sigma^2)]$, $S(x)$ is calculated as $S(x)\propto1-\exp[-\sigma^2(x/c)^2/2]$, which is identical to the time-forward HOM interferogram.


\begin{figure}[t]
  \centering
  \includegraphics[width=8cm]{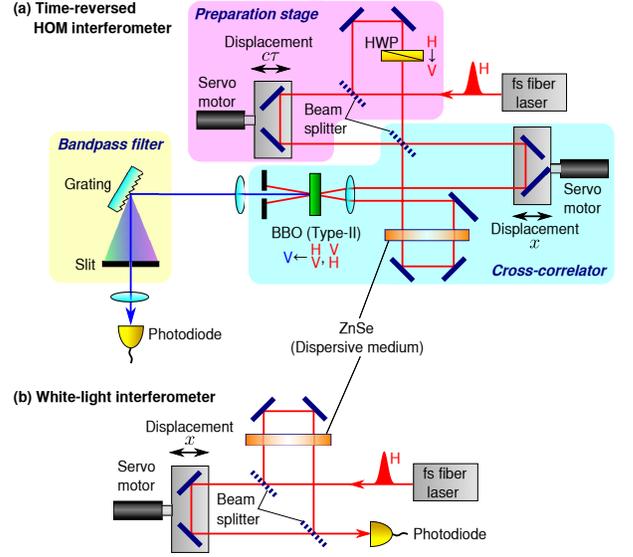}
\caption{(color online).
(a) Experimental setup of the time-reversed HOM interferometer. 
Pairs of orthogonally polarized pulses with time difference $\tau$ are prepared in the preparation stage composed of a beam splitter, a half-wave plate (HWP), and a delay mirror moved by a DC servo motor.
The cross-correlator is composed of a beam splitter, a delay mirror moved by a DC servo motor, and a BBO crystal for Type-II non-collinear SFG.
The bandpass filter passing narrow-band (0.43\,nm) light is composed of a grating and a slit. 
(b) Experimental setup of the white-light interferometer.
We used the same light source and dispersive medium as for the time-reversed HOM interferometer.} 
\label{exp}
  \end{figure}

\begin{figure*}[t]
  \centering
  \includegraphics[width=14cm]{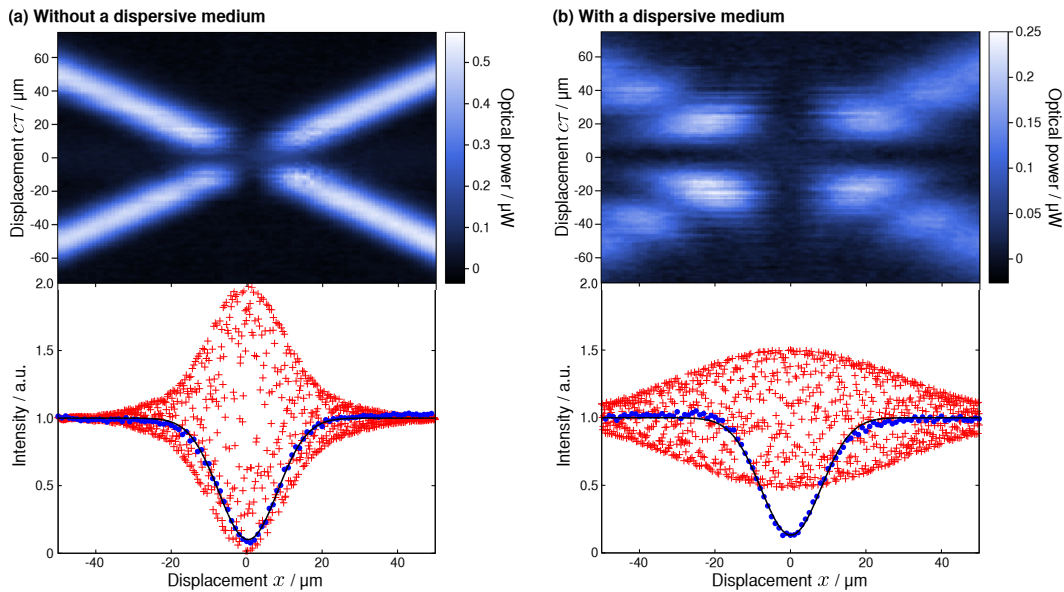}
\caption{(color online).
Experimental results of the time-reversed HOM interferometer and the white-light interferometer (a) without and (b) with a dispersive medium.
The upper color maps show the intensity distribution for all $x$ and $c\tau$. (Note that the color scales in (a) and (b) differ.)
In the lower plots, the blue circles show the time-reversed HOM interferograms, derived by integrating the upper data over all $c\tau$.
The red crosses show the white-light interferograms for comparison.
} 
\label{map}
  \end{figure*}

We next experimentally generate dispersion-insensitive HOM interferograms with the time-reversed HOM interferometer.
Our experimental setup is shown in Fig.~\ref{exp}(a).
A femtosecond fiber laser (center wavelength 782\,nm, pulse duration 74.5\,fs FWHM, average power 54\,mW) was used to create transform-limited pulsed light with horizontal polarization.
In the preparation stage, each pulse is divided into a pair of pulses at the first non-polarizing beam splitter. One of the pair experiences a relative path difference $c\tau$ and the other is rotated into vertical polarization by a half-wave plate (HWP).
The pair of pulses is introduced into the cross-correlator with time difference $\tau$.
The cross-correlator introduces a relative path difference $x$ in one of its arms and includes a dispersive medium in the other.
We use a 5-mm-thick zinc selenide (ZnSe) plate as a dispersive medium. 
After propagating in each arm, the pulses are focused into a 1-mm-length $\upbeta$-barium borate (BBO) crystal for type-II non-collinear SFG.
The sum-frequency light is filtered into a 0.43-nm bandwidth around 391\,nm by a 1,200-lines/mm aluminum-coated diffraction grating followed by a slit, and then detected by a GaP photodiode equipped with an amplifier.
We measured the intensity $I\sub{r}(x,\tau)$ for various displacements $c\tau$ and integrated $I\sub{r}(x,\tau)$ for all displacements $c\tau$ to obtain the HOM interferograms.
We also observed white-light interferograms for comparison, using an ordinary Mach--Zehnder interferometer with the same light source shown in Fig.~\ref{exp}(b).

The experimental results without and with a dispersive medium are shown in Figs.~\ref{map} (a) and (b), respectively.
The upper color maps show the intensity distribution versus displacement $x$ and $c\tau$ in the time-reversed HOM interferometer.
Both $x$ and $c\tau$ are spaced at intervals of 1\,$\upmu$m.
The lower graphs show time-reversed HOM interferograms (blue circles) and white-light interferograms (red crosses).
The time-reversed HOM interferograms are derived by vertically integrating the upper graphs over all $c\tau$. 
Without a dispersive medium, we observed 18.3$\pm$0.2\,$\upmu$m FWHM for the time-reversed HOM dip and 24.63$\pm$0.04\,$\upmu$m FWHM for the white-light interference pattern; the former had a narrower width by a factor of 1.35.
This factor is slightly smaller than the theoretical value $\sqrt{2}$ for input light with a Gaussian spectrum \cite{PhysRevA.88.043845} due to the lost bandwidth in SFG.
With the dispersive medium, we observe 18.6$\pm$0.2\,$\upmu$m FWHM for the time-reversal HOM dip and 69.5$\pm$0.3\,$\upmu$m FWHM for the white-light interference pattern. 
The width of the time-reversed HOM dip remains essentially unchanged owing to automatic dispersion cancellation, whereas that of the white-light interference pattern is clearly increased by a factor of 1.8. 

We note that our results achieve high-visibility HOM interferograms with high signal conversion efficiency. 
The measured visibilities of the time-reversed HOM dips are 89.9$\pm$0.7\% without dispersion and 86.8$\pm$0.8\% with dispersion. 
From the maximum signal power 0.57\,$\upmu$W and the light source power 54\,mW, the maximum signal conversion efficiency is estimated to be $1.1\times 10^{-5}$.
This value is about one order of magnitude greater than that for CPI, $1.6\times 10^{-6}$, which is estimated from the previous experiment \cite{kaltenbaek2008quantum}.
These improvements are due to the simplicity of our setup, which does not require the generation of chirped laser pulses.


In conclusion, we have proposed a classical optical interferometry reproducing dispersion-insensitive HOM interferograms,
employing a systematic method based on the time-reversal symmetry of quantum mechanics. 
We have also experimentally demonstrated automatic dispersion cancellation using this interferometry. 
Our results exhibit high-visibility interference dips with high signal conversion efficiency,
demonstrating that the method is appropriate for metrological applications.
Furthermore, the time-reversal method illustrated here can be applied to the development of a new classical interferometric technique reproducing various quantum phenomena.

This research is supported by JSPS KAKENHI Grant Number 22109004 and 25287101.



\end{document}